\documentclass[11pt]{article}
\usepackage{rotating}

\usepackage{lscape}

\usepackage{graphicx}
\begin{document}
\title{\bf Relationship between the Elemental Abundances and the Kinematics
of Galactic-Field RR~Lyrae Stars}

\author{{V.\,A.~Marsakov, M.\,L.~Gozha, V.\,V.~Koval'}\\
%EndAName
{Southern Federal University, Rostov-on-Don, Russia}\\
{e-mail:  marsakov@sfedu.ru, gozha\_marina@mail.ru, vvkoval@sfedu.ru}}
\date{accepted \ 2018, Astronomy Reports, Vol. 62, No. 1, pp. 50-62}

\maketitle

\begin {abstract}

Abstract--Data of our compiled catalog containing the positions, 
velocities, and metallicities of 415 RR~Lyrae variable stars and 
the relative abundances [el/Fe] of 12~elements for 101 RR~Lyrae stars,
including four $\alpha$~elements (Mg, Ca, Si, and Ti), are used to 
study the relationships between the chemical and spatial--kinematic
properties of these stars. In general, the dependences of the relative 
abundances of $\alpha$~elements on metallicity and velocity for the 
RR~Lyrae stars are approximately the same as those for field dwarfs. 
Despite the usual claim that these stars are old, among them are 
representatives of the thin disk, which is the youngest subsystem 
of the Galaxy. Attention is called to the problem of low-metallicity 
RR~Lyrae stars. Most RR~Lyrae stars that have the kinematic properties 
of thick disk stars have metallicities ${\rm [Fe/H]} < -1.0$ and 
high ratios [$\alpha$/Fe]$ \approx 0.4$, whereas 
only about 10\,\% of field dwarfs belonging to the so-called 
``low-metallicity tail'' have this chemical composition. At the same 
time, there is a sharp change in [$\alpha$/Fe] in RR~Lyrae stars belonging
just to the thick disk, providing evidence for a long period of 
formation of this subsystem. The chemical compositions of 
SDSS J1707+58, V455 Oph, MACHO 176.18833.411, V456 Ser, and BPS CS 
30339--046 do not correspond to their kinematics. While
the first three of these stars belong to the halo, according to their 
kinematics, the last two belong to the thick disk. It is proposed that 
they are all most likely extragalactic, but the possible appearance 
of some of them in the solar neighborhood as a result of the 
gravitational action of the bar on field stars cannot be ruled out.

\end{abstract}

{{\bf Key words:} RR~Lyrae stars, chemical composition, kinematics, 
Galaxy (Milky Way)}.

\maketitle

%%%%%%INTRODUCTION%%%%%%%%%%
\section{Introduction}

RR Lyrae variables populate the horizontal branch
of the Hertzsprung--Russell diagram. They are considered
to be typical Population II objects according
to the classification of Baade, and are among the
oldest stars of the Galaxy. These high-luminosity
stars are easily identified due to their short-period
brightness variability, they are visible to large distances,
and their absolute magnitudes can be fairly
reliably calculated using their metallicities. For this
reason, field RR Lyraes are often used in studies of
the structure and evolution of the young Galaxy. The
heliocentric distances of field RR Lyrae stars and the
tangential components of their velocities are continuously
being refined. For example, Dambis et\,al. \cite{1}
recently used proper motions, visual magnitudes, and
two infrared magnitudes from the UCAC4 together
with the statistical-parallax method to refine the zero
points of the dependences of the absolute magnitudes
on metallicity and the variability period, and provided
corresponding calibration relations. The distances
and velocities of four hundred Galactic RR~Lyrae
stars obtained from the data of \cite{1} are currently the
most accurate and homogeneous.

It is not possible to estimate ages of individual
RR~Lyrae stars using theoretical evolutionary tracks,
since their positions in the Hertzsprung--Russell diagram
are essentially independent of age. Studies of
their chemical compositions can help to some extent
in putting order on the chronology of the formation
of these stars. Various elements are synthesized in
thermonuclear fusion reactions in stars with different
masses which evolve on different timescales and
eject these elements into the interstellar medium at
different epochs. In particular, $\alpha$~elements, 
rapid neutron capture elements, and a small
number of iron-peak elements are ejected by massive
Type II supernovae a few tens of millions of years
after the formation of their progenitors. On the other
hand, most iron-group elements are formed in Type Ia
supernovae, which occur approximately a billion years
after the formation of their progenitors (see, for example,
\cite{2}). Therefore, in a closed star–gas system, the
ratios [$\alpha$/Fe] and [r/Fe] for stars formed from interstellar
material enriched in SN II products will begin
to decrease steadily in time approximately a billion
years after a burst of star formation. Thus, the relative
abundances of these elements will become statistical
indicators of the ages of these stars. Despite the
complex processes taking place in the interiors of
RR~Lyrae stars, the chemical compositions of their
atmospheres usually reflect the chemical composition
of the environment in which they were formed (see, for
instance, \cite{3} and references therein).

Traditionally, the population of field RR~Lyrae stars
is represented as components of two subsystems of
the Galaxy -- the halo and thick disk (see, for example,
\cite{1}). However, even a glance at the chemical and
kinematic parameters of individual RR~Lyrae stars
finds among them both stars with essentially solar
chemical compositions and velocities and those with
retrograde orbits, low metallicities and relative elemental
abundances significantly different from the
solar values. Therefore, the aim of this study was to
compile a catalog of spectroscopic determinations of
the metallicities, the relative elemental abundances,
particular of $\alpha$~elements, and the spatial velocity components
for as much RR~Lyrae stars in the Galactic
field as possible, with the aim of using the data of this
catalog to identify regular relationships between these
parameters for RR~Lyrae stars belonging to different
Galactic subsystems.

%%%%%% Data %%%%%%%%%%

\section {INPUT DATA}

For our multi-faceted study, we took the catalog
\cite{1} as the main source of spatial and kinematic data.
This catalog contains the variability periods, metallicities
calculated using the Preston index, proper motions,
and radial velocities of 392 RR~Lyrae stars. The 
distances to the stars were calculated by applying one
of three equivalent period-–metallicity--luminosity
calibration relations presented in the catalog, based
on the infrared magnitude $K_{s}$. We then used these
distances to calculate the Cartesian coordinates ($x$,
$y$, $z$), and spatial velocty components ($V_R$, $V_{\Theta}$, $V_Z$)
in cylindrical coordinates, corrected for the solar
motion relative to the Local Standard of Rest (local
centroid, LSR). The components of the solar motion
relative to the LSR were taken to be $(U,V,W)_\odot  = (11.1, 12.24, 7.25)$
~km/s \cite{4}, the solar Galactocentric
distance to be 8.0~kpc, and the rotation velocity of
the local centroid to be 220~km/s. The differences
between the distances calculated using the three
calibration relations given in \cite{1}, based on different
magnitudes, are no greater than 10\,\%. Given this
uncertainty and the uncertainties in the radial velocities
and proper motions for each RR~Lyrae star in the
catalog, the mean uncertainty in the spatial velocity
is 16~km/s. We found information for calculating
the Cartesian coordinatse and velocities of 15 RR
~Lyrae stars with known chemical compositions that
were not included in \cite{1} from several sources in the
literature. References to the sources of the distances,
proper motions, and radial velocities are listed in our
compiled catalog (see below).

Atmospheric elemental abundances provide information
on the chemical evolution of the interstellar
matter from which they were formed, and thus make it
possible to trace the chemical evolution of the Galaxy.
Unfortunately, the exposures required for obtaining
high-quality spectra with high signal-to-noise ratios
for distant RR~Lyrae stars are limited by their short
pulsation periods and the high amplitudes of their
radial-velocity variations in their envelopes. Nevertheless,
a large number of elemental abundances for
field RR~Lyrae stars determined from high-quality
spectra obtained on fairly large telescopes have been
published recently. These data have shown that the
strongest and most symmetric spectral lines are obtained
close to the minimum brightness (maximum
radius) of the variable star, when the atmosphere is
stationary (see, for example \cite{5,6}). The temperature,
gravity and microturbulence velocities in the stellar
atmosphere vary synchronously with phase, while the
derived elemental abundances (apart from those for
silicon and barium) are essentially insensitive to the
phase \cite{7,8}. Therefore, in most cases, the spectra
used to determine the abundances were obtained during
quiescent phases close to 0.35.

We collected spectroscopic determinations of the
relative abundances of 13 elements in 101 field RR
Lyrae stars from the literature (25 papers published
in 1995--2017); references are provided in the catalog.
We selected data on the abundances of $\alpha$
~elements (O, Mg, Si, Ca, Ti), iron-peak elements
(Fe), neutron-capture elements (Y, Ba, La, Zr, Eu),
and elements with an odd number of protons (Na,
Al). The elemental abundances were determined using
high-resolution spectra. The analysis of most
of the spectra was carried out in the approximation
of local thermodynamic equilibrium (LTE), but 
non-LTE effects were taken into account in a number
of studies. In the vast majority of cases, Kurucz
model atmospheres \cite{9} were used. About a hundred
RR~Lyrae stars had two values of [Fe/H]. One of
the metallicities in \cite{1} was calculated based on the
Preston index and calibrated using the Zinn--West
scale, while the other was an average over spectroscopic
determinations of iron abundances collected
from the literature. Examination of these data showed
that the metallicity scales do not display systematic
differences, and the metallicities for almost all the RR
~Lyrae stars are the same within the uncertainties.

The data from various studies for which the solar
abundances used were indicated were reduced to
the solar composition recommended in \cite{10}. In our
case, only such a correction was possible. In a small
number of studies, the adopted scale for the solar
composition was not indicated. In these cases, the
relative abundances were not corrected. For 59 RR
~Lyrae stars, abundances were determined in a single
study only. For 42 RR~Lyrae stars, the abundance
of a given element was determined more than once
(up to eight times); in such cases, we computed
the weighted mean values with coefficients inversely
proportional to the uncertainties claimed in those
studies. The quoted uncertainties in several studies
exceeded 0.3 dex. Unfortunately, in two cases, these
were the only determinations available. If the relative
uncertainties were not specified, we assumed them to
be 0.2 for studies published before 2000 and 0.1 for
later papers. The mean uncertainties for each element
proved to be in the range $\varepsilon \rm{[el/Fe]} = (0.11 - 0.18)$, 
and the mean value for all elements was 
$\langle \varepsilon \rm{[el/Fe]} \rangle = 0.14$.

To verify the external convergence of the determinations
of the abundance of each element, we constructed
distributions of the deviations of the relative
abundances derived in a given study for a specified
star from the weighted mean values. All these histograms
are described well by normal laws, consistent
with the uncertainties being random. The dispersions
of the distributions for the different elements are in
the range $\sigma \rm{[el/Fe]} = (0.06 - 0.17)$, while the mean
value for all the elements is 
$\langle\sigma\rm{[el/Fe]}\rangle = 0.11$. This
means that the external convergence of the relative
abundance determinations is even slightly lower than
the uncertainties claimed in the initial studies. Analysis
of the external convergence demonstrated an absence
of systematic offsets. Although the number of
overlapping determinations for most of the elements
is not very large, we believe that our compilation of
abundances of elements produced in various nuclear
fusion processes in field RR~Lyrae stars can be used
to study the chronology of their formation and the
evolution of the Galaxy.

As a result, we compiled a catalog of the kinematic
and chemical parameters of 415 field RR~Lyrae
variables. For 407, we provide distances and coordinates,
for 401 velocity components, and for 101
relative elemental abundances. A fragment of the
catalog is shown in the Table{1}\footnote{The Table is available in 
full only in electronic form.}. 
The columns of
the Table present (1) the name of the star, (2)--(3)
the Galactic coordinates ($l$, $b$), (4) the fundamental
pulsation period (for RRc stars, this was calculated
using the formula log $logP_{F}=logP+0.127$ \cite{11}), (5)
the heliocentric distance $d$, (6)--(8) coordinates ($x$,
$y$, $z$) in a right-handed orthogonal system, (9) the
Galactocentric distance $R_{G}$ in kpc, (10)--(12) the
calculated spatial velocity components relative to the
LSR ($U$, $V$, $W$)$_{LSR}$, (13)--(15) the velocity components
in cylindrical coordinates ($V_{R}$, $V_{\Theta}$, $V_{Z}$), where
$V_{R}$ is directed toward the Galactic anticenter, $V_{\Theta}$ in
the direction of the Galactic rotation, and $V_{Z}$ toward
the North Galactic pole, (16) references for the distance,
proper motion, and velocity, (17) the values of
[Fe/H]$_{D}$ from \cite{1}, (18)--(30) the relative abundances
of iron [Fe/H] and twelve other elements ([el/Fe])
we have found, and (31) references to the sources
of the abundances. A key for the references for the
kinematic data, [Fe/H], and [el/Fe] is given in the
electronic catalog. The stars are listed alphabetically
in the catalog in order of the constellation names or
the names given in the original sources of data.

We used the data of Bensby et al. \cite{12}, which contains
metallicities, relative abundances of $\alpha$~elements,
and spatial-velocity components for 714 F\,--G\,field
dwarfs, for comparison.

\section {SEPARATION OF RR LYRAE STARS
OVER THE GALACTIC SUBSYSTEMS} 

There is no single necessary and sufficient criterion
that can be used to assign each star to a particular
subsystem of the Galaxy with absolute confidence.
Old objects of the Galaxy, such as globular
clusters, subdwarfs and RR~Lyrae stars, are usually
grouped into two subsystems---the thick disk and the
halo. In this case, it is convenient to distinguish
the subsystems according to metallicity, since this
is an indicator of age albeit somewhat rough. The
metallicity distributions of these objects show a dip
close to $\rm{[Fe/H]}\approx -1.0$, like globular clusters (see,
for example, [13, Fig. 1]) or a break of the metallicity
function, like field RR~Lyrae stars (see, for example,
\cite{14} and Fig. 1 in that paper). Such a feature
in the metallicity distributions of globular clusters
and subdwarfs, as well as red dwarfs and giants with
$\rm{[Fe/H]} < -0.5$, which were earlier considered to be
typical representatives of the spherical subsystem of
the Galaxy, was first noted in \cite{15, 16}. Since the
purpose of our current study is to investigate the
chemical compositions of RR~Lyrae stars in different
subsystems, we distinguished the subsystems based
on the kinematic parameters. For this, we used the
methodology suggested in \cite{17}, where the probabilities
for field stars to belong to the thin disk, thick disk,
and halo are computed using their spatial-velocity
components relative to the local centroid and the
dispersions of these components in each subsystem.
This method assumes that the spatial-velocity components
of the stars in each subsystem obey normal
distributions.

Application of this technique showed that, out of
the 401 stars in our catalog with known velocities,
56 most likely belong the thin disk subsystem, 122
to the thick disk, and 223 to the halo. According
to our current understanding, the halo consists of
two unrelated subsystems—the intrinsic halo and an
accreted halo (see, for example, \cite{13,14,18}). The
objects in the intrinsic halo are genetically related to
objects in the younger subsystems of the Galaxy---
the thin and thick disks formed of matter from the
same protogalactic cloud. Some globular clusters
and individual stars making up the accreted halo were
captured by the Galaxy from dwarf satellite galaxies
that were disrupted by Galactic tidal forces, and
are formed of matter that went through a different
chemical evolution history. For simplicity, we distinguished
such stars here only via their retrograde
motion around the Galactic center, since we believe
that this is the most characteristic signature of an
extragalactic origin. More than half of the halo RR
Lyrae stars in our catalog (139 objects) have $V_{\Theta} <
0$ km/s. Since the subsystems do not have sharp
boundaries and are embedded in one another, the
velocity dispersions and other kinematic parameters
characterizing the properties of the Galactic subsystems,
which are included in formulas defining the
separation of stars over the subsystems, are determined
only approximately. To minimize ambiguity in
assigning a particular star to a particular subsystem,
it is recommended to consider a star to belong to a
given subsystem if the probability that it belongs to
another subsystem is at least a factor of two lower
(for reliability, sometimes even a factor of ten) \cite{17}.
According to this criterion, we cannot assign a number
of stars in the sample to a subsystem, even when
kinematic data are available. Such stars are usually
referred to as intermediate stars. Since we are interested
in statistical regularities in the subsystems,
which can be identified only using sufficiently large
samples of objects, we assigned all of the stars to
subsystems, taking a greater probability of belonging
to one particular subsystem to be sufficient.

Figure 1a shows the distributions of F\,--G\, dwarfs
from \cite{12} and of field RR~Lyrae stars of the selected
subsystems in the Toomre diagram, that is, 
``$V_{LSR} - (U^{2}_{LSR}+W^{2}_{LSR})^{0.5}$''. 
Generally, the distributions of
both types of objects are approximately the same.
However, in the transition to older subsystems, the
fraction of RR~Lyrae stars becomes appreciably larger
than the fraction of typically younger main-sequence
stars. It is noteworthy that, according to their kinematic
parameters, RR~Lyrae stars, which have traditionally
been regarded as typical representatives of
the very old subsystems of the Galaxy, are present in
the fairly young thin-disk subsystem. (Note that, in
[8], using a probabilistic kinematic criterion similar to
ours, but with other kinematic data, 10 out of 23 high metallicity
field RR~Lyrae stars with spectroscopic
elemental abundances, as in our study, were defined
to be thin-disk objects). At the same time, a significant
fraction of RR~Lyrae stars are even overtaking
the local centroid in its rotation around the Galactic
center. In our samples, such stars comprise more
than 40\,\% of thin disk field RR~Lyrae stars, while they
constitute only one-third of field dwarfs. To examination
the reliability of assigning the RR~Lyrae stars
to the thin disk, we applied a recurrent procedure and
set the parameters in the formulas used to calculate
the probabilities equal to the values we obtained for
the RR~Lyrae stars in the different subsystems. In
this case, all the velocity dispersions in the subsystems
increased slightly, but the fraction of stars in
the thin-disk subsystem drastically decreased. The
recalculation only redistributed the allocations of a
few RR~Lyrae stars, whose kinematics indicate they
are in the transition zone between the thin and thick
disks. In particular, among RR~Lyrae stars with
spectroscopic elemental abundances, only PH Peg
and BPS CS 30339--046 changed subsystems.

Figure 1b shows the distribution of RR~Lyrae stars
in a plot of metallicity versus the distance from the
Galactic plane (z). The RR~Lyrae stars of the thick
disk are located mainly at large distances from the
Galactic plane and, unlike the thin-disk RR~Lyrae
stars, do not show an increased concentration toward
this plane. As a result, the scale height of
the subsystem formally defined by these stars, $Z_{0} = 1.1 \pm 0.1$~
kpc, is larger than the scale height defined
by field dwarfs (see, for example, \cite{19}, where 
$Z_{0} = 0.6 \pm 0.1$~kpc). This is due to observational selection
effects caused by the increase in absorption with
approach toward the Galactic plane. As a result,
the relative number of distant RR~Lyrae stars with
small $z$ values is underestimated. Let us now see
how justified assignment to the thick-disk subsystem
based on metallicity is for field RR~Lyrae stars.

The histograms in Fig.~1d, where the metallicity
functions of field RR~Lyrae stars and field dwarfs of
the thick disk are compared, clearly show that there
is an significant difference between them. Unexpectedly,
the overwhelming majority of the thick-disk
RR~Lyrae stars (83\,\%) have metallicities $\rm{[Fe/H]} < -1.0$. 
The thick-disk dwarfs also include some
low-metallicity stars.the so-called ``low-metallicity
tail''. However, according to many studies, the
fraction of such stars barely exceeds 10\,\% (see, for
example, \cite{20}). The metallicity functions in Fig.~1d
show that the lowest metallicity objects of both types
reach approximately the same metallicity, $\rm{[Fe/H]}\approx -2.0$. 
However, among the more metal-rich stars,
the metallicity function of the RR~Lyrae stars with
the kinematics of the thick disk reaches only half
the corresponding distribution for the field dwarfs,
and ends at approximately [Fe/H]~$\approx -0.5$: there are
no more metal-rich field RR~Lyrae stars with thick-disk
kinematics, whereas the nature of field dwarfs
with essentially solar metallicities and thick-disk
kinematics has been very actively discussed (see, for
instance, \cite{12}). A similar shift is observed for the
metallicity function of the thin-disk RR~Lyrae stars.
In the halo, the ranges of the metallicity distributions
of the two types of objects are similar, but their
maxima are also separated, though only slightly (the
histograms are not present in order to economize
space).

Figure~1e presents the metallicity as a function
of the azimuthal velocity ($V_{\Theta}$) for the same objects.
Small symbols show RR~Lyrae stars with metallicities
from \cite{1} determined using PrestonЃfs index,
and large symbols show objects with spectroscopic
metallicities. The two slanted almost parallel dashed
lines show an approximate separation of the stars into
Galactic subsystems by eye, without computing the
probability that they belong to a particular subsystem.
The separation corresponding to these lines satisfactorily
correlates with the separation according to our
probability criterion, and the lines themselves pass
through the regions occupied by stars with uncertain
membership. Since, as we noted above, there is no
unambiguous criterion for separating the stars into
subsystems, it is impossible to unconditionally reject
any of the approaches. Moreover, the velocities of
some of the stars could well be distorted by radial
migration. The diagram also shows that several RR
~Lyrae stars with thick-disk kinematics turned out
to be less metal-rich than the F\,--G\, dwarfs with the
lowest metallicities in the same subsystems. Most of
them got fell into the zone of uncertain separation,
and may actually belong to the thick disk. (Liu
et\,al. \cite{8} found velocity components of DH Peg ---
one of the lowest metallicity objects among the RR
~Lyrae stars they classified (or indeed among those
classified by us)---assigned to the thin disk in another
source, according to their kinematics; it was
subsequently reclassified even not as a thick-disk,
but as a halo star). Figure 1e shows that neither
in the thin disk, nor in the halo are there obvious
dependences between the azimuthal velocities of the
stars and their metallicities. However, a progressive
decrease in the upper limit of the metallicity with
decreasing azimuthal velocity ($V_{\Theta}$) is observed in the
thick-disk stars. Note that the two lowest-metallicity
RR~Lyrae stars with spectroscopic determinations of
[Fe/H] (V456 Ser and BPS CS 30339--046), which
we placed in the thick disk with a high probability,
are missing from the catalog \cite{1} with homogeneous
spatial and kinematic parameters, and their distances
and velocities were taken from other sources. In
particular, these data for the latter star were taken
from \cite{21}, which is devoted exclusively to this star.
The iron abundance in these RR~Lyrae stars is much
lower than those in the other thick disk stars. One of
these stars has a rotational velocity around the Galactic
center exceeding the solar value. They are also far
from the other stars in the other diagrams (see below).
It is very possible that these RR~Lyrae stars, which
have velocities typical for thick-disk objects, actually
entered the Galaxy from disrupted satellite galaxies,
like the stars of the Arcturus moving group (see \cite{19},
\cite{22}). According to its spatial-velocity components,
one of these RR~Lyrae stars (V456 Ser) may even
belong to this stellar stream.

Figure 1e also shows that almost all RR~Lyrae
stars with retrograde rotations have metallicities
$\rm{[Fe/H]} < -1.0$, and these RR~Lyrae stars constitute
a majority in the halo subgroup. A similar ratio of
the numbers of genuine Galactic objects and accreted
objects is obtained for the field stars \cite{23}. However,
RR~Lyrae stars with the kinematics of the accreted
halo also include some very metals-rich objects.
For example, V455 Oph has a spectroscopic metal
abundance exceeding the solar value. However,
with the metallicity determined using the Preston
index [Fe/H]$ = -1.42$, which is consistent with the
distance \cite{1}, it falls into the center of the distribution
of halo RR~Lyrae stars in a ``$V_{\Theta}$ -- [Fe/H]'' diagram.
(This is one of the three RR~Lyrae stars of our sample
for which the difference between the metallicities
determined using the two methods exceeds than
0.5 dex). Note, if we recalculate its heliocentric
distance taking the value of $\rm{[Fe/H]}_{Sp} = 0.19$ used by us, the
azimuthal velocity remains zero within the uncertainties,
although it formally becomes positive. In other
words, its orbit remains essentially perpendicular to
the Galactic plane. One of the RR~Lyrae stars from
our catalog that is most remote and has among the
lowest metallicities---SDSS J170733.93+585059.7
(henceforth, SDSS J1707+58)---also has a retrograde
orbit.

The the full residual velocity of the star relative to
the local centroid is sometimes used to separate stars
into subsystems. Figure~1f shows the distributions
of stars in a plot of [Fe/H] vs. $V_{res}$. Unlike the
azimuthal velocity, the residual velocity has all spatial
components. The dependence of the metallicity on the
velocity in the thick disk stands out more clearly in
this diagram. At the same time, it is clear that this
kinematic parameter can also be used as a statistical
indicator of membership of a star to one or another
subsystem. The two vertical dashed lines in this
diagram approximately separate stars of the thick and
thin disks ($V_{res} \approx 80$~km/s) and those of the thick
disk and halo ($V_{res} \approx 200$~km/s). Note that if the
latter value of the residual velocity is exceeded, many
objects with retrograde orbits appear, as is clearly
seen in the diagram. This diagram also shows that the
two low-metallicity RR~Lyrae stars noted above have
residual velocities almost in the middle of the range,
typical for the thick disk. By the way, the prototype of
the population, RR~Lyrae, is also in the thick disk; it
is located almost in the Galactic plane, and its vertical
velocity component is zero within the uncertainties.
Although its azimuthal velocity is quite typical for this
subsystem, $V_{\Theta} = 112$ km/s, its residual velocity is
very high, $V_{res} = 257$~km/s. More than half a century
ago, based on its kinematic and photometric parameters,
RR~Lyrae was included in the moving group of
fast halo stars Groombridge 1830 \cite{24}. The last two
diagrams also show that the line $\rm{[Fe/H]} = -1.0$ collects
into the group of metal-rich field dwarfs nearly
all stars of both disk subsystems, whereas it collects a
large fraction of the field RR~Lyrae stars with the thin disk
kinematics, and only a small fraction with the
thick-disk kinematics. It turns out that the chemical
and kinematic criteria for separating the subsystems
of RR~Lyrae stars are not unambiguous.

\section {RELATIONSHIPS BETWEEN THE RELATIVE
ABUNDANCES OF $\alpha$~ELEMENTS AND OTHER PARAMETERS}

Figure~2 shows the dependences of the relative
abundances of magnesium, silicon, calcium and
titanium on [Fe/H] for F\,--G\, field dwarfs and field RR
Lyrae stars. For both types of object, the sequences
of relative abundances of magnesium and calcium
are fairly narrow, and practically coincide when
$\rm{[Fe/H]} > -1.0$, while RR~Lyrae stars have slightly
higher [Mg/Fe] and [Ca/Fe] values than the field
dwarfs at lower metallicities. In the lower-metallicity
range, most of the relative elemental abundances
were taken from \cite{25},\cite{26}. Similar excesses at low
metallicities are show by two other $\alpha$~elements ---
silicon and titanium. However, there is a very large
scatter in the relative abundance of silicon, and all the
RR~Lyrae stars have high metallicity, but significantly
lower [Ti/Fe], than the field dwarfs. The relative
abundances $\rm{[Fe/H]} > -1.0$ were main determined
in \cite{8}; however, they are consistent with the results
of other studies for this range. We emphasize that
the noted deviations of the relative abundances of the
listed $\alpha$~elements are not a consequence of differences
in the methods used to determine them in the dwarfs
and RR~Lyrae stars, since these deviations were
also noted in the original studies, and their nature
remained unexplained, although it was emphasized
there that they cannot be interpreted in terms of the
chemical evolution of the Galaxy \cite{8,25}. Thus, the
systematic deviations of the sequences of $\alpha$~elements
for the RR~Lyrae stars from the analogous sequences
for the field dwarfs could well be due to distortions
associated with the influence of the continually
changing physical conditions in the atmospheres of
variable stars. Further, to minimize any artificially
created patterns, we will consider the behavior of the
dependences of the relative abundances of $\alpha$~elements
on the metallicity and velocity averaged only over the
two $\alpha$~elements magnesium and calcium, for which
the systematic deviations are within the uncertainties
in the abundances.

Figure~3a shows the dependences of [Mg,Ca/Fe]
ratios averaged in this way on the metallicity for field
dwarfs and RR~Lyrae stars. The two types of objects
show fairly close sequences. For comparison,
Fig.~3b presents the same dependences averaged over
all four $\alpha$ elements which show good agreement with
Fig.~3a. Unfortunately, all four $\alpha$ elements were
simultaneously determined for a smaller number of
RR~Lyrae stars. At the same time, the sequences
became narrower than in Fig.~3a for both types of
objects, and a gap between the thin and thick disks
near $\rm{[\alpha/Fe]} \approx 0.16$ became distinctly visible for the
dwarfs. While the systematic differences from the
dwarfs are more sharply manifested for the RR~Lyrae
stars, and the RR~Lyrae stars as a whole are lower
in the high-metallicity range, the RR~Lyrae stars as
a whole are higher compared to the field dwarfs in
the low-metallicity range. As was noted above, the
reason for this effect remains unclear.

It follows from Fig.~3a that the dependence of
the [Mg, Ca/Fe] ratios on [Fe/H] in thin-disk RR
Lyrae stars is in good agreement with its behavior
for the field dwarfs. An exception is the RR~Lyrae
star DH Peg, which has a lower metallicity than the
lowest metallicity field dwarfs in this subsystem. In
addition, the relative abundance of $\alpha$~elements in this
star are somewhat higher compared to the remaining
thin-disk RR~Lyrae stars. Note that DH Peg is
classified as a halo object in \cite{8} based on its velocity
components, and its chemical composition is then
consistent with those of either halo stars or thick-disk
stars. TV Lib, in the metallicity
range characteristic of most thin-disk stars, also
have high [$\alpha$/Fe] values. In fact, the abundances of
this star were determined in the only study, and
it proved impossible to reduce the data to a single
solar composition. According to their positions in
Figs.~3a, b, DH Peg and TV Lib are more naturally
classified as thick-disk stars. The RR~Lyrae star
KP Cyg, which has the lowest relative abundance in
our sample, $\rm{[Mg,Ca/Fe]} = -0.18$, also belongs to
the thin disk, according ot its kinematics. As can
be seen from Figs.~2b and~2d, the abundances of
two other $\alpha$~elements (Si and Ti) in this star are, on
the contrary, enhanced. The study \cite{27} is devoted to
an analysis of the chemical composition of this star,
which has a very high metal abundance and abnormally
high abundances of carbon and nitrogen in its
atmosphere. It is proposed in \cite{27} this star, as well
as UY CrB, are in reality not long-period RR~Lyrae
stars, but instead short-period CWB-type cepheids.
In other words, their presence in a list of RR~Lyrae
stars is questionable. The unusual nature of these
stars is also confirmed by their anomalous positions in
some of our diagrams (the abundances of $\alpha$~elements
were not determined for UY CrB, but it has a large
distance from the Galactic plane in Fig.~1b).

The largest systematic differences between the
field RR~Lyrae stars and dwarfs are observed for the
stars with thick-disk kinematics. One of the differences
has already been noted: instead of a sparsely
populated low-metallicity tail, as is present for the
thick-disk dwarfs, low-metallicity stars dominate
among the thick-disk RR~Lyrae stars. Figure~3a
also shows that the thick-disk dwarfs show a distinct
break in the [Fe/H] -- [$\alpha$/Fe] diagram near 
$\rm{[Fe/H]} \approx -0.5$ (see also \cite{28}). 
However, the break in Figs.~3,a,b
for the RR~Lyrae stars is located near $\rm{[Fe/H]} \approx -1.0$.
The presence of a small break at this location for
thick-disk field dwarfs was also noted earlier in \cite{9}.
However, this conclusion has been questioned in
some other studies \cite{29}. A reduction in the relative
abundance of $\alpha$~elements in field RR~Lyrae stars,
starting from $\rm{[Fe/H]} \approx -1.0$, was also noted in \cite{8}.
However, we note that the conclusion that a break is
present at this metallicity for the RR~Lyrae stars is
not statistically significant, due to the small number
of objects. We can also see that one of the two thick disk
RR~Lyrae stars with uncharacteristically low
metallicities for this subsystem---BPS CS 30339--046---
has a very low relative abundance of $\alpha$~elements,
compared to the average value for stars of
the same metallicity. This difference is appreciably
higher than the uncertainties stated in the studies
where the abundances were calculated. However,
the RR~Lyrae star V456 Ser, which satisfies the
kinematic criterion for membership in the Arcturus
stream, falls in the middle of the general sequence
for the low-metallicity stars in Fig.~3a, just like the
detected field stars in this stream \cite{12,19}. This
provides further evidence for its membership in the
Arcturus stream. However, it is positioned somewhat
lower in Fig.~3b, where the sequence for the averaged
values of all four investigated $\alpha$~elements is shown,
due to the fairly low silicon abundance of this star.
Although the kinematics of another RR~Lyrae star,
TY Gru, correspond to the thick disk, it has an
abnormally low abundance of heavy elements for this
subsystem, and is located very far from the Galactic
plane ($z  = -4.2$~kpc).

RR~Lyrae stars and dwarfs exhibiting halo kinematics
behave approximately the same. We expect
an appreciable spread in the investigated relations
among the field stars with retrograde orbits, since
the atmospheric chemical compositions of stars that
were presumably formed in various satellite galaxies
may be different from the composition of genetically
related Galactic objects with similar metallicities. Indeed,
in one of the lowest metallicity stars of the
sample, SDSS J1707+58, the abundance of $\alpha$~elements 
turned out to be abnormally high compared
to all the field stars. Note that the abundances of 
$\alpha$~elements in SDSS J1707+58 were averaged over the
results of two studies, and the values of [el/Fe] for the
corresponding elements were very close.

The RR~Lyrae star V455 Oph, which has retrograde
rotation and is located in the strip of thin-disk
stars, also occupies an anomalous position in
Figs.~3,a,b. Figures~3,c,d show relations between
the averaged relative abundances of two $\alpha$~elements
and the kinematic parameters of the investigated
stars. Since the velocities, like the metallicities,
are statistical indicators of age, it is not surprising
that the dependences of [$\alpha$/Fe] on these parameters
are somewhat similar. However, there are also
peculiarities. For example, two halo stars (MACHO
176.18833.411 and AO Peg) with metallicities
and relative abundances of $\alpha$~elements that are
consistent with both the thick disk and halo (see
Fig.~3a) have velocities around the Galactic disk
that are much higher than the solar value (Fig.~3c).
AO Peg has a total residual velocity of 409~km/s, and
MACHO 176.18833.411 a total velocity of 485~km/s
(Fig.~3d). In \cite{30}, which is devoted to the latter object,
taking into account its position near the Galactic
center and the shape of its orbit, it is concluded that
the star was most likely thrown out of the Galactic
center and has its origin in the low-metallicity tail of
the Galactic bulge.

The RR~Lyrae star that is the most metal-poor
and simultaneously has the highest [$\alpha$/Fe] value
and exhibits the kinematics of the accreted halo,
SDSS J1707+58, also has an exceptionally high
negative azimuthal velocity (see Fig.~1e and Fig.~3c).
The accreted halo RR~Lyrae star V455 Oph has a
virtually zero azimuthal velocity ($V_{\Theta}$), high $V_{res}$, and
a very low [$\alpha$/Fe] that differs from the values
for the remaining halo stars beyond the uncertainties.
Two RR~Lyrae stars (TV Lib and DH Peg) with
high [$\alpha$/Fe] values, although classified by us as thin
disk stars based on both their azimuthal and residual
velocities, could equally be attributed to the thick
disk.

\section {DISCUSSION} 

Thus, our analysis of the chemical and kinematic
properties of field RR Lyrae variable stars has shown
that such stars are present in all four subsystems
of the Galaxy we have distinguished: the thin and
thick disks, and the intrinsic and accreted Galactic
halos. Thus, contrary to traditional ideas, our sample
of field RR~Lyrae stars includes objects with kinematics
and chemical compositions typical of stars of
the thin disk. Modern estimates of the age of the
thin disk are $<9$~Gyr (see, for instance, \cite{12}). Hence,
contrary to traditional picture (e.g., \cite{31}), RR~Lyrae
stars include both old ($>10$~Gyr) objects and younger
stars. On the other hand, the thin-disk RR~Lyrae
stars include objects with lower metallicities and with
higher relative abundances of $\alpha$~elements than those
for field dwarfs in this subsystem. For the absolute
majority of thin-disk RR~Lyrae stars, the [$\alpha$/Fe] ratios
are close to solar, as is also true for F\,--G\,~dwarfs of this
subsystem.

Strictly speaking, RR~Lyrae stars with metallicities
characteristic of the thin disk should not exist,
since the horizontal branch for such stars is located
in the region of the red-giant clump, alongside the
instability strip, and these stars should not be variable.
However, according to our results, the kinematics
and chemical compositions of these stars with
high probability indicate that they belong precisely to
the thin disk. This means that the reason for this
discrepancy should be sought in their classification
as variable stars. This list should include at least
two more high-metallicity RR~Lyrae stars from other
subsystems --- AA Aql and V455 Oph. We already
noted above that the RR Lyrae star from our list
with the highest metallicity and the longest period,
KP Cyg, is probably a classical Cepheid with a very
short period \cite{27}. It may be that the metal-rich stars
we are discussing do not pulsate in the fundamental
mode, but in an overtone, in which case all of them
could actually be Cepheids. Examples of Cepheids
with periods less than a day include V1334 Cyg,
V1726 Cyg, and V1154 Cyg from the catalog \cite{32}.
It is unlikely that they are $\delta$~Sct variables, which
are Population I stars, since the variability periods of
$\delta$~Sct stars are shorter than those of RR~Lyrae stars.
In any case, their verification requires a thorough
investigation of every such a variable.

Most of the RR~Lyrae stars with thick-disk kinematics
turned out to have [Fe/H] values that would
usually be considered to be in the low-metallicity tail
of thick-disk field stars. This can be explained by the
fact that, being older than most dwarfs, they trace the
chemical composition of the interstellar medium in
the initial stages of the formation of this subsystem.
The break in the dependence of [$\alpha$/Fe] on [Fe/H]
indicates that the epoch of Type Ia supernova had
begun in the star-gas system; i.e., about 1~Gyr had
passed since the onset of star formation. Apparently,
the first SNe Ia began to explode when the metallicity
of the interstellar medium in the Galaxy reached
$\rm{[Fe/H]} \approx -1.0$, and only when higher metallicities
were reached did SNe Ia begin to explode en masse.
The long duration of the evolution of the thick-disk
subsystem is testified to by the systematic trends in
both the metallicity and the relative abundance of ѓї
elements with changes in the kinematic parameters
within a given subsystem, which is clearly visible in
Figs.~1e,f and Figs.~3c,d. These dependences support
the hypothesis of a prolonged formation for the thick
disk in the process of the collapse of the protogalactic
cloud.

The two lowest metallicity RR~Lyrae stars of this
subsystem (BPS CS 30339--046 and V456 Ser)
demonstrate chemical compositions that differ from
those of the other RR~Lyrae stars of the subsystem,
beyond the uncertainties. This suggests that they
have an extragalactic origin, similarly to the stars of
the well-known Arcturus stream. The spatial velocity
of the RR~Lyrae star V456 Ser suggets that it may
well also belong to this stream. The very low value of
$\rm{[Mg/Ca]} = -0.3$ and the high relative abundance of
the fast-neutron-capture element $\rm{[Eu/Fe]} = 1.0$ for
V456 Ser \cite{25} also supports an extragalactic origin
for this star. These are the most extreme values for
the RR~Lyrae stars of our sample (see our catalog).
Both these quantities indicate a low mass of the Type
II supernovae that enriched the interstellar material
from which this star was formed. According to our
current understanding, the yield of primary $\alpha$~elements
(such as magnesium) compared to secondary
$\alpha$~elements (such as calcium) decreases with the
pre-supernova mass, while virtually all europium is
formed in the r-process, which occurs during the
least massive SNe II explosions, with masses of
$(8-10)M_{\odot}$ (see, e.g., \cite{33}). Low-mass supernovae
explode with higher probability in dwarf low-mass
galaxies \cite{34}. Note, however, that some authors prefer
the hypothesis that the Arcturus stream, to which this
RR~Lyrae star probably belongs, formed as a result
of the gravitational perturbation of field stars by the
Galactic bar (see \cite{12} and references therein).

In contrast to V456 Ser, BPS CS 30339--046
shows a very low relative abundance of $\alpha$~elements
and a very low metallicity. This could arise if this star
formed in a dwarf galaxy, where the rate of star formation
was so low that SNe Ia began to explode when
the interstellar medium was still not very enriched in
iron from SNe II. However, these possibilities require
additional research.

Virtually all the RR~Lyrae stars with halo kinematics
have relative abundances of $\alpha$~elements that
correspond to their metallicity. This applies fully to
the RR~Lyrae stars of the intrinsic Galactic halo,
that is, to stars that are genetically related to the
protogalactic cloud. The RR~Lyrae stars of the
accreted halo, to which we have attributed all stars
with retrograde orbits, have a somewhat larger spread
of the relative abundances of some $\alpha$ elements for
the given metallicity. This circumstance requires
a more careful determination of the atmospheric
chemical compositions of these stars, which have
often been determined in one study only. Like
another low-metallicity RR~Lyrae star of the thick
disk, TY Gru, SDSS J1707+58 is considered to be
a low-metallicity star with an enhanced abundance
of carbon and $s$ elements. It has been suggested
that this RR Lyrae star is a binary and that its
companion is losing mass, being in the asymptotic
branch giant stage (see \cite{21} and references therein).
At the same time, there are doubts as to how such
a star could complete its evolution in the red giant
branch without experiencing the consequences of
Roche lobe overflow. However, it turns out that it also
has an enhanced relative abundance of $\alpha$ elements.
The very high ratio [$\alpha$/Fe] in SDSS J1707+58 may
be more easily understood if its protostellar cloud was
enriched by the ejecta of a very massive SN II, whose
explosion provoked star formation in this cloud, also
accelerating it.

We also consider the chemical composition of
V455 Oph to be unique. While this star is located
in the general sequence of thin-disk stars in
the [Fe/H]--[$\alpha$/Fe] diagram and has almost solar
elemental abundances, it has a retrograde orbit
that is nearly perpendicular to the Galactic plane.
This high abundance of metals in combination with
such a high orbit could hardly arise in a low-mass
dwarf satellite galaxy. According to the numerical
model for the hierarchical formation of the Galactic
halo \cite{35}, only a low-mass satellite galaxy could be
destroyed by the tidal forces of the Galaxy while
still distant from the Galactic plane, that is, in a
perpendicular orbit. It is also possible that this star,
like MACHO 176.18833.411, was ejected away from
the Galactic bulge. Its current distance from the
center of the Galaxy is very large, $R_{G}\approx 6.8$~kpc,
but it has a high velocity in the direction toward the
center, 163~km/s. Clarifying this situation requires
investigating the trajectory of its orbit. We emphasize
again that additional research is required to verify all
our conclusions concerning individual stars.

We plan to investigate the behavior of the relative
abundances of other elements in field RR~Lyrae
stars from our catalog and the relationship between
the chemical compositions and the parameters of the
Galactic orbits of these stars in a future paper.

%%%%%%ACKNOWLEDGMENTS%%%%%%%%%%
\section*{ACKNOWLEDGMENTS}

We thank Alexei Rastorguev, one of the authors
of the catalog of spatial and kinematic parameters of
RR~Lyrae stars, for calculations of the distances to
the RR~Lyrae stars using the method they developed
(these distances were not given in the published version
of the catalog). This work was supported by
the Ministry of Education and Science of the Russian
Federation (state contracts 3.5602.2017/BCh and
3.858.2017/4.6).

\renewcommand{\refname}{REFERENCES}

\newpage

\begin{figure*}
\centering
\includegraphics[angle=0,width=0.99\textwidth,clip]{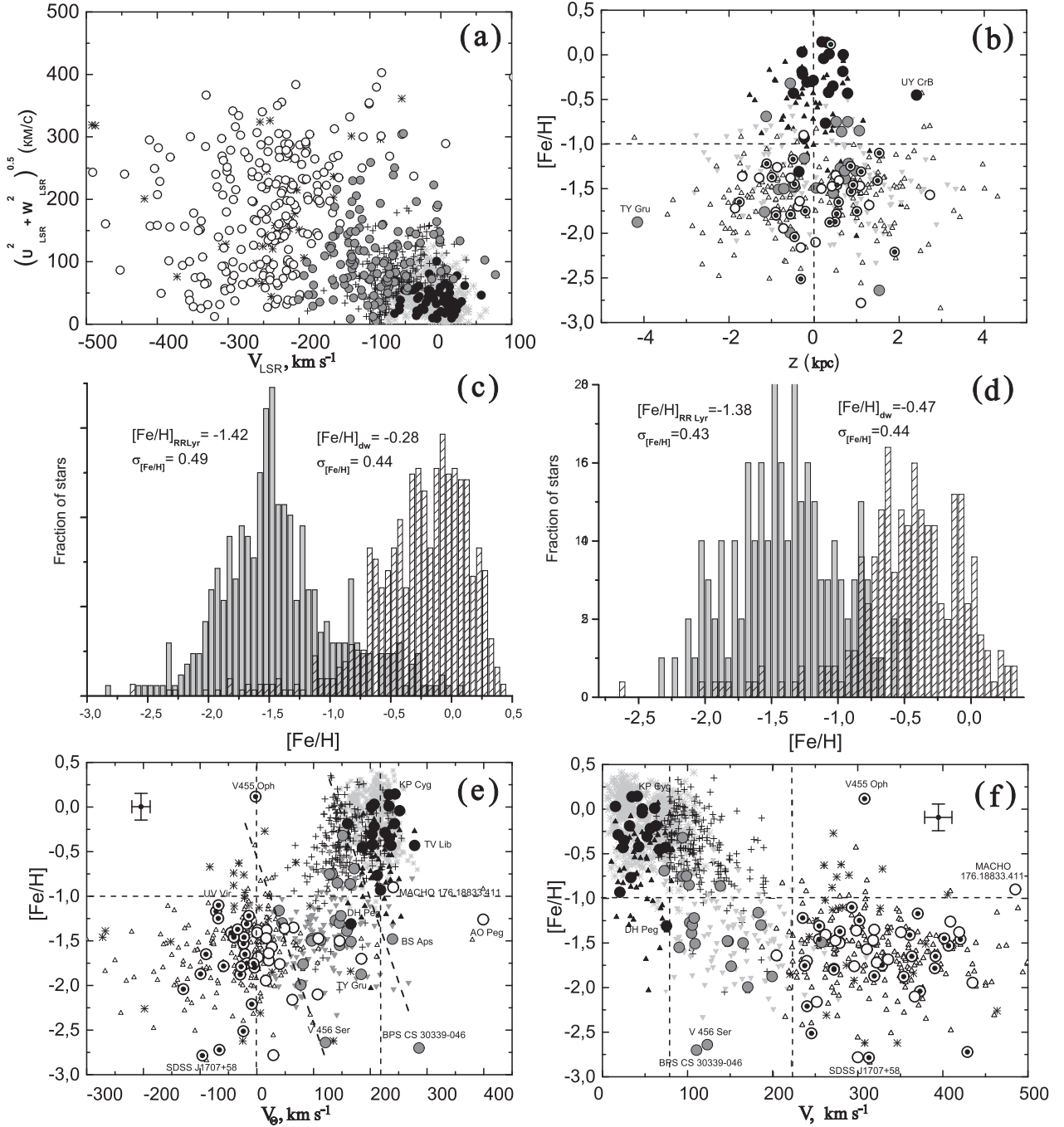}
\caption{(a) Toomre diagram, (b) dependence of metallicity on 
         distance from the Galactic plane, metallicity distributions 
         for (c) all stars of the sample and (d) thick disk stars only, 
         and dependence of metallicity on (e) the rotational velocity 
         about the Galactic center and (f) the total residual velocity 
         for 714 F\,-G\, dwarfs and 401 field RR~Lyrae stars. Thin disk
         field dwarfs are shown by light gray asterisks, thick disk 
         field dwarfs by gray pluses, and halo field dwarfs by black 
         asterisks. Field RR~Lyrae stars with metallicities from \cite{1}
         are shown by small triangles, and those with spectroscopic
         metallicities by large circles. Dark circles show RR Lyrae 
         stars in the thin disk, thick disk, and halo are shown by 
         filled circles, gray circles, and hollow circles, respectively;
         a dark dot inside a large circle indicates a star displaying 
         retrograde rotation. The slanted (e) and vertical (f) dashed 
         lines show the conventional separation of stars into different 
         subsystems in the diagram. The dashed horizontal lines are 
         drawn at [Fe/H]~$= -1.0$ (b, e, f). The vertical dashed lines 
         are drawn through the values $V_{\Theta} = 0$~ and ~220 km/s 
         (e). RR~Lyrae stars with spectroscopic metallicities or 
         velocities that strongly deviate from the mean velocities for 
         the corresponding subsystems are named (d, e). In the 
         metallicity distributions, RR~Lyrae stars are plotted 
         in gray, and dwarfs are shown by slanted shading. The numbers 
         indicate the average metallicities and their dispersions (c, d).}
\label{fig1}
\end{figure*}

\newpage

\begin{figure*}
\centering
\includegraphics[angle=0,width=0.99\textwidth,clip]{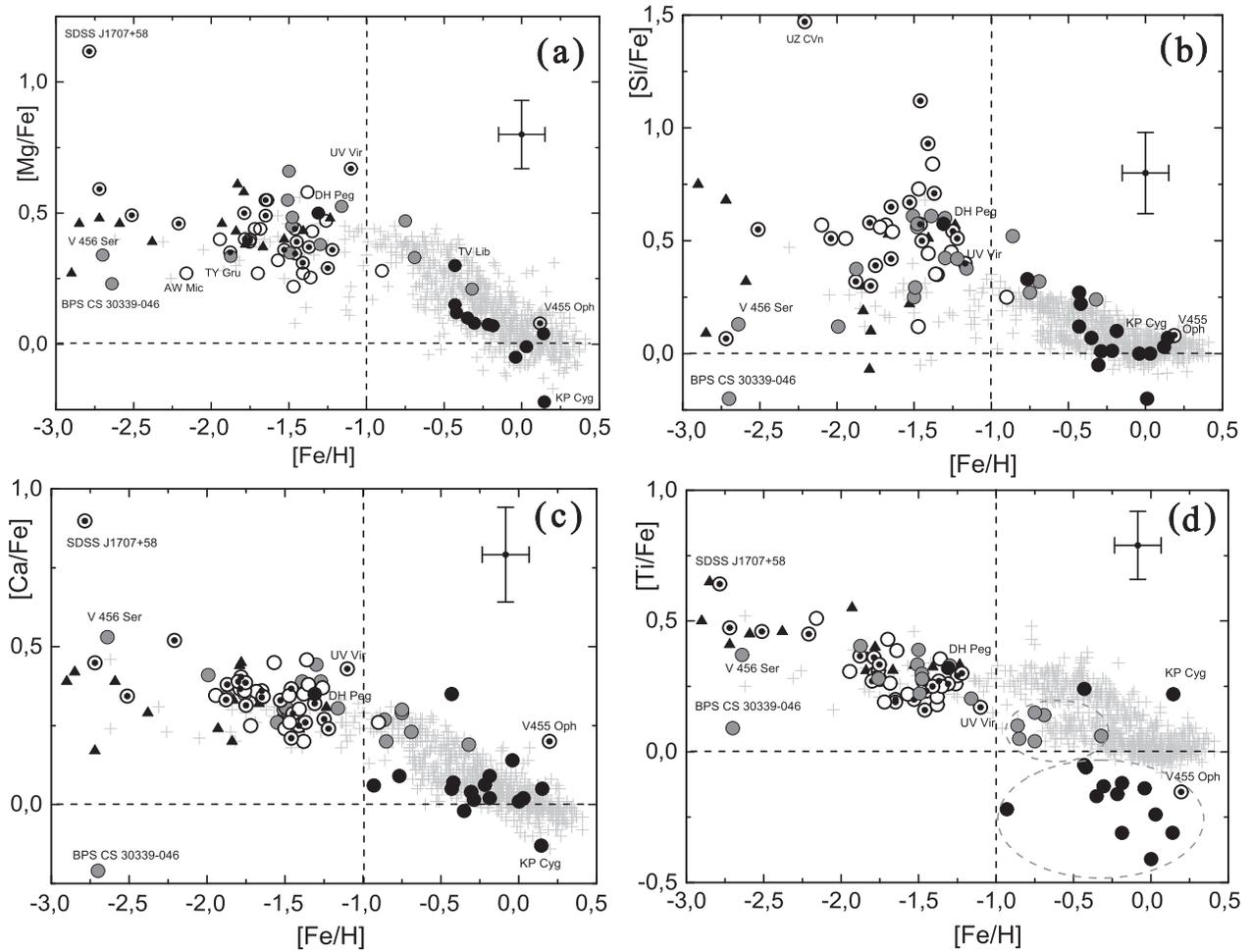}
\caption{Dependence of relative abundances of (a) magnesium, (b) 
         silicon, (c) calcium, and (d) titanium on the metallicity 
         for field dwarfs and RR~Lyrae stars. The notation is the 
         same as in Fig.~1.}
\label{fig2}
\end{figure*}

\newpage

\begin{figure*}
\centering
\includegraphics[angle=0,width=0.95\textwidth,clip]{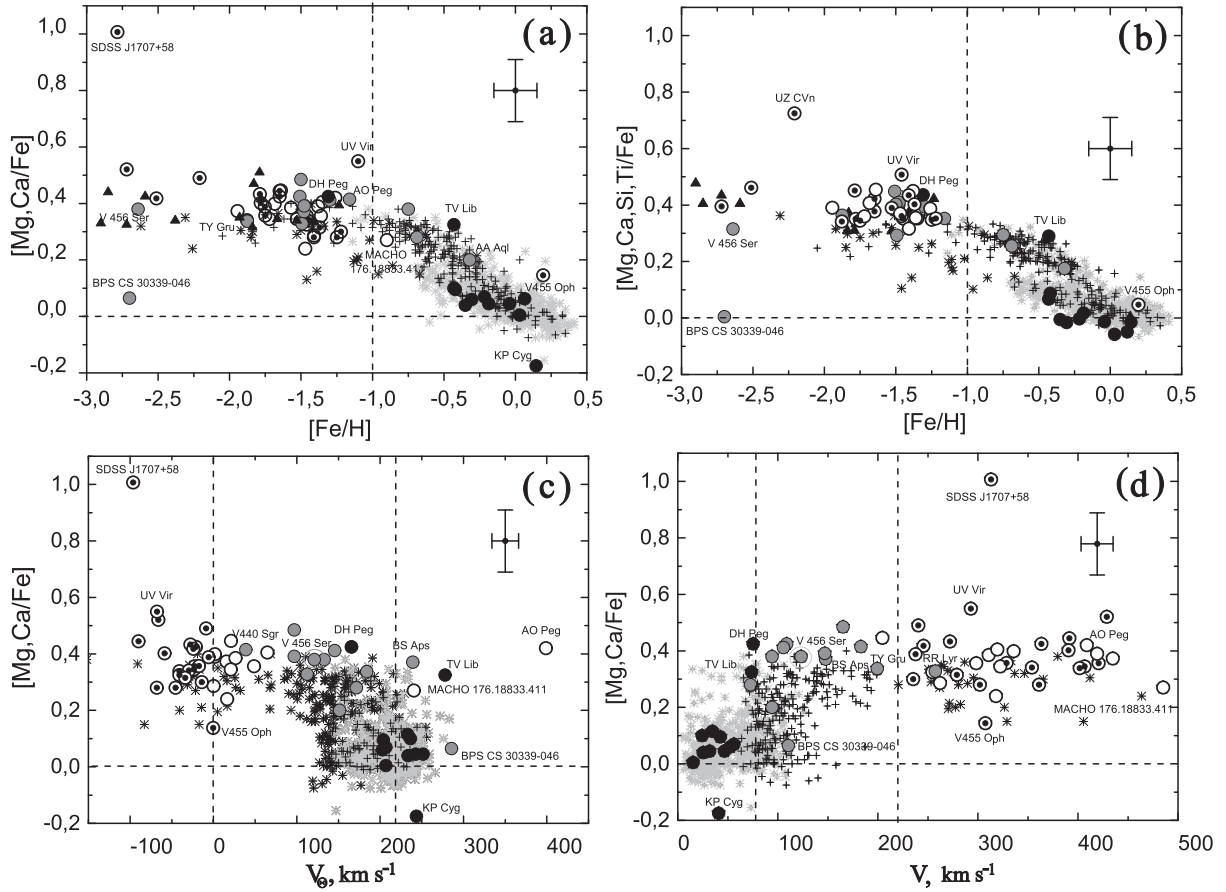}
\caption{Dependence of the relative abundances on the metallicity. 
         Panel (a) shows the abundances averaged over two 
         $\alpha$~elements (Mg and Ca), and Panel (b) the abundances 
         averaged over four $\alpha$~elements (Mg, Ca, Si, and Ti). 
         Panels (c) and (d) show the 
         dependence of [Mg,Ca/Fe] on the rotational velocity around 
         the Galactic center and the total residual velocity for field 
         dwarfs and RR~Lyrae stars. The notation is the same as in Fig.~1.}
\label{fig3}
\end{figure*}

\newpage
\clearpage

\newpage

\begin{landscape}

\begin{table}[t!]

\caption{%
Kinematic parameters and relative elemental abundances in the field 
RR Lyrae stars (a fragment of the catalog)}
\bigskip
%\label{tproto}
\begin{center}
\begin{tabular}{c|c|c|c|c|c|c|c|c|c|c}
\hline \hline

\multicolumn{1}{c|}{\parbox{1.6cm}{Name}}&
\multicolumn{1}{c|}{\parbox{1.2cm}{$l$,$^{\circ}$}}&
\multicolumn{1}{c|}{\parbox{1.2cm}{$b$,$ ^{\circ}$}}&
\multicolumn{1}{c|}{\parbox{1.8cm}{$P_F$, day}}&
\multicolumn{1}{c|}{\parbox{1.2cm}{$d$, kpc}}&
\multicolumn{1}{c|}{\parbox{1.2cm}{$x$, kpc}}&
\multicolumn{1}{c|}{\parbox{1.2cm}{$y$, kpc}}&
\multicolumn{1}{c|}{\parbox{1.2cm}{$z$, kpc}}&
\multicolumn{1}{c|}{\parbox{1.6cm}{$R_{G}$, kpc}}&
\multicolumn{1}{c|}{\parbox{1.2cm}{$U_{LSR}$, km/s}}&
\multicolumn{1}{c}{\parbox{1.2cm}{$V_{LSR}$, km/s}}\\

\hline
   1  &  2     & 3      &  4   & 5   &  6   & 7    &  8   &9    &10   &11\\
\hline
SW And&115.7250&-33.0825&0.4423&0.495&-0.180& 0.374&-0.270&8.193&  50.7&-15.2\\
CI And&134.9313&-17.6169&0.4848&1.597&-1.075& 1.078&-0.483&9.152&  -9.2&19.1\\
DR And&126.1660&-28.5670&0.5328&2.072&-1.074& 1.469&-0.991&9.245&-160.2&-228.3\\
BS Aps&317.8839&-15.1835&0.5826&1.738& 1.244&-1.125&-0.455&6.864&-146.0&-1.9\\
KP Cyg& 77.2592&  5.0360&0.8360&2.147& 0.472& 2.086& 0.188&7.814&  27.1&25.2\\
\hline

\end{tabular}
\end{center}

%\label{tproto}
\begin{center}
\begin{tabular}{c|c|c|c|c|c|c|c|c|c|c}

\hline \hline

\multicolumn{1}{c|}{\parbox{1.6cm}{Name}}&
\multicolumn{1}{c|}{\parbox{1.2cm}{$W_{LSR}$, km/s}}&
\multicolumn{1}{c|}{\parbox{1.2cm}{$V_{R}$, km/s}}&
\multicolumn{1}{c|}{\parbox{1.2cm}{$V_{\Theta}$, km/s}}&
\multicolumn{1}{c|}{\parbox{1.2cm}{$V_{Z}$, km/s}}&
\multicolumn{1}{c|}{\parbox{2.6cm}{Ссылка на $d$, pmRA, pmDE, RV}}&
\multicolumn{1}{c|}{\parbox{1.2cm}{$[Fe/H]_{D}$}}&
\multicolumn{1}{c|}{\parbox{1.2cm}{[Fe/H]}}&
\multicolumn{1}{c|}{\parbox{1.2cm}{[O/Fe]}}&
\multicolumn{1}{c|}{\parbox{1.2cm}{[Na/Fe]}}&
\multicolumn{1}{c}{\parbox{1.2cm}{[Mg/Fe]}}\\

\hline
  1   &  12 & 13  &14   & 15  &16&17  & 18  &19  &20  &21   \\
\hline
SW And&-18.6&-41.3&206.9&-18.6&7&-0.38&-0.22&0.22&0.09& 0.08\\
CI And&-11.8& 37.3&236.3&-11.8&7&-0.83&-0.43&   -&0.01& 0.15\\
DR And&  5.7&156.8&-33.8&  5.7&7&-1.48&-1.37&   -&   -& 0.37\\
BS Aps& 22.2&108.2&239.1& 22.2&7&-1.33&-1.48&   -&0.15& 0.45\\
KP Cyg& 17.3& 39.4&243.6& 17.3&13, 18&& 0.15&0.06&0.31&-0.22\\
\hline
\end{tabular}
\end{center}

%\label{tproto}
\begin{center}
\begin{tabular}{c|c|c|c|c|c|c|c|c|c|c}

\hline \hline

\multicolumn{1}{c|}{\parbox{1.6cm}{Name}}&
\multicolumn{1}{c|}{\parbox{1.2cm}{[Al/Fe]}}&
\multicolumn{1}{c|}{\parbox{1.2cm}{[Si/Fe]}}&
\multicolumn{1}{c|}{\parbox{1.2cm}{[Ca/Fe]}}&
\multicolumn{1}{c|}{\parbox{1.2cm}{[Ti/Fe]}}&
\multicolumn{1}{c|}{\parbox{1.2cm}{[Y/Fe]}}&
\multicolumn{1}{c|}{\parbox{1.2cm}{[Zr/Fe]}}&
\multicolumn{1}{c|}{\parbox{1.2cm}{[Ba/Fe]}}&
\multicolumn{1}{c|}{\parbox{1.2cm}{[La/Fe]}}&
\multicolumn{1}{c|}{\parbox{1.2cm}{[Eu/Fe]}}&
\multicolumn{1}{c}{\parbox{2.4cm}{Ссылка на [Fe/H], [el/Fe]}}\\

\hline
1     &  22 &23  & 24  & 25  & 26  & 27 & 28  & 29  &30  &31   \\
\hline
SW And&-0.07&0.01& 0.06&-0.16&-0.53&   -&-0.26&    -&   -&5,10,21,22,30,31\\
CI And&    -&0.12& 0.05&-0.05&    -&   -&-0.21&    -&   -&22\\
DR And&    -&0.71& 0.26& 0.27&    -&   -&    -&    -&   -&23\\
BS Aps& 0.59&0.56& 0.29& 0.30& 0.04&0.57& 0.04&-0.03&0.14&4, 12\\
KP Cyg& 0.25&0.07&-0.13& 0.22&-0.02&   -&-0.35&    -&   -&2, 30\\

\hline
\end{tabular}
\end{center}

\end{table}
\end{landscape}

%\end{landscape}

\end{document}